\documentclass[conference]{IEEEtran}
\IEEEoverridecommandlockouts
\usepackage[toc,acronym]{glossaries}
\usepackage{adjustbox}
\usepackage{tikz}
\usepackage{listings}
\usetikzlibrary{arrows,shadows,backgrounds,calc,decorations.pathreplacing,decorations.pathmorphing,positioning,
	automata}
\usepackage{subcaption}
\makeglossaries

\newacronym{AI}{AI}{Artificial Intelligence}
\newacronym{ANN}{ANN}{Artificial Neural Network}
\newacronym{ALU}{ALU}{Arithmetical and Logical Unit}
\newacronym{CPU}{CPU}{Central Processing Unit}
\newacronym{FPGA}{FPGA}{Field Programmable Gate Array}
\newacronym{GPU}{GPU}{Graphic Processing Unit}
\newacronym{HW}{HW}{hardware}
\newacronym{ICCB}{ICCB}{Inter-Core Communication Block}
\newacronym{ISA}{ISA}{Instruction Set Architecture}
\newacronym{I/O}{I/O}{Input/Output}
\newacronym{LAN}{LAN}{Local Area Network}
\newacronym{MC}{MC}{Multi-Core and/or Many-Core}
\newacronym{MLP}{MLP}{Memory Level Parallelism}
\newacronym{OoO}{OoO}{Out-of-Order}
\newacronym{OS}{OS}{operating system}
\newacronym{PD}{PD}{Propagation Delay}
\newacronym{QT}{QT}{Quasi-Thread}
\newacronym{PU}{PU}{Processing Unit}
\newacronym{SPA}{SPA}{Single Processor Approach}
\newacronym{SW}{SW}{software}
\newacronym{HPL}{HPL}{High Performance Linpack}
\newacronym{HPCG}{HPCG}{High Performance Conjugate Gradients}

\newacronym{EMPA}{EMPA}{Explicitly Many-Processor Approach}
\newacronym{EPE}{EPE}{EMPA Processing Element}
\newacronym{EME}{EME}{EMPA Morphing Element}
\newacronym{ECE}{ECE}{EMPA Communicating Element}
\newacronym{EICB}{EICB}{EMPA Inter-Core Block}
\newacronym{ESME}{ESME}{EMPA Storage Manager Element}

\definecolor{webgreen}{rgb}{0,.5,0}
\definecolor{webbrown}{rgb}{.6,0,0}
\definecolor{webyellow}{rgb}{0.98,0.92,0.73}
\definecolor{webgray}{rgb}{.753,.753,.753}
\definecolor{webblue}{rgb}{0,0,.8}
\definecolor{webgreen}{rgb}{0, 0.5, 0} % less intense green
\definecolor{webred}{rgb}{0.5, 0, 0}   % less intense red

\usepackage{cite}
\usepackage{amsmath,amssymb,amsfonts}
\usepackage{graphicx}
\usepackage{textcomp}
\usepackage{xcolor}
\def\BibTeX{{\rm B\kern-.05em{\sc i\kern-.025em b}\kern-.08em
		T\kern-.1667em\lower.7ex\hbox{E}\kern-.125emX}}
\begin{document}
	
	\title{Introducing temporal logic to computer science*\\
		\thanks{Submitted to the 5th IEEE International Conference on Rebooting Computing. Project no.136496  has been implemented with the support provided from the National Research, Development and Innovation Fund of Hungary, financed under the K funding scheme.
		}
	}
	
	\author{\IEEEauthorblockN{J\'anos V\'egh}
	\IEEEauthorblockA{
	\textit{Kalim\'anos BT}\\
	Debrecen, Hungary \\
	Vegh.Janos@gmail.com~ORCID:~0000-0002-3247-7810}
	%\and
	%\and
	%\and
	%\and
	%\and
	}
	
	\maketitle
	
	\begin{abstract}
		Computing uses the general operating model, that data is delivered to the input of the processing element, and after some processing time, the resulting data is delivered to some store. In his classic publication, von Neumann
		assumed that "in the actually intended vacuum tube interpretation \dots the conduction [transfer] times may indeed by neglected".
		Since that time, this paradigm is followed, independently whether 
		the data are delivered via direct wiring or using packages over networks, and whether the processing element is digital/analog electronic component, FPGA, biological/technical neuron or quantum processor, although von Neumann warned: "the conduction [transfer] times \dots can be longer than the synaptic delays [processing time], hence our  procedure of neglecting them aside of $\tau$ [processing time] would be \textit{unsound}."
		He has not foreseen the present technological conditions, however, he made sure "It would take an ultra high frequency device -- $\tau\approx10^{-8}$ seconds or less -- to vitiate this argument". That is, for the today's GHz processors (and especially, using them in extreme-size supercomputers or in systems mimicking biological operation), the classic paradigm is "\textit{unsound}". 
		
		In computing, everything is about time, from producing faster 
		processors, interconnections and memories, to
		providing supercomputers with higher \textit{payload} performance or
		reducing the weeks-long training time of Artificial Neural Networks.
		Despite this, \textit{in the commonly used classical computing paradigm,
			the time has no explicit role,} in the sense that only
		the logical dependence of the operations is considered, but not that \textit{the physical implementation of the components of computing systems
			converts that logical dependence to temporal dependence}.
		
%		The case of computing can be paralleled with the case of science. The classic science could explain nearly everything
%		around us, assuming instant interaction between its objects.
%		It became evident about a century ago, however, 
%		that the interaction speed was not infinitely large,
%		and introducing  an insurmountable 
%		interaction speed (the speed of light), led to the birth of "modern sciences",
%		such as relativistic and quantum physics.
%		The more general reason was that natural objects behave differently
%		at extremely high speeds or extremely low amounts of energy.
%		Similar \textit{unusual behavior can be observed at other extreme conditions}, such as big (or small) masses, sizes, densities, etc.
		
		In computing, the ratio of the processing to the transfer time has reversed in the past decades. Moore's observation was correct only to
		the density of gates inside processors, but the physical size of computer components persisted in staying
		in the cm-range, and the operating clock speed has grown to million-fold higher.
%		These differences also mean that the computing states change
%		so quickly, that time-dependence of the operations must be considered:
%		the idea of instant change throughout a circuit is not valid anymore.
		\textit{A modern, time-aware computing paradigm is needed.} 
		
		Due to the incremental nature of computing,
		the stealthy effect was noticed, but was handled inadequately, via introducing clock domains and distribution trees.
		To compensate their temporal behavior,
		about half of the power consumption of processors is required (and another half is needed for cooling).
		In supercomputers, only about one percent of the power consumption of the vast amount of processors is
		used for payload calculation for real-life tasks.
		When attempting to create biology-mimicking computing systems,
		without using the principles of the biological evolution, 
		the payload performance stalls about a couple of dozens of
		networked processors.
		Not considering the temporal behavior of the implementation of the components
		led to stalling in all computing fields, from single/many-core processor performance to payload performance of supercomputing and ANN systems.
%		The effect was topped with accepting the fake idea of "weak scaling", for scaling distributed systems.
		
		The idea proposed here, is, to introduce a \textit{temporal logic}, i.e., unlike in the classical computing,
		the value of a logical expression depends on WHEN and WHERE
		it is evaluated.
		This solution keeps the solid mathematical background of computing science; only the logical functions will depend 
		on those extra arguments.
		
		The parallel with the modern science also suggests the method
		of introducing such a change. The Minkowski-transformation 
		assumes that there exists a limiting speed, and the world
		shall be described by space-time coordinates.
%		instead of space coordinates only. Einstein assumed that that limiting
%		speed was the speed of light.
About a century ago, the distance
		was the quantity that could be reliably measured,
		so \textit{the time coordinate was transformed into space coordinate}.
		
		In computing systems, the reverse of the Minkowski transform shall be followed: we need to use a \textit{time-space coordinate system},
		i.e.,\textit{ we propose to transform the space coordinates to time coordinates}
		(the space coordinates are distances measured along their path of wiring).
%		In that coordinate system, Einstein's classical hypothetical experiment can be reproduced. The only new idea is that switching the light on, has a \textit{processing time}.
		We propose that the (apparent) \textit{ total execution time comprises both transfer time and processing time}.
		The fact that the transfer time matters is evident in biological systems. However, their cyclic operation and that they can
		modulate their conduction velocity, mostly covers that evidence.
		
		The idea of considering their temporal behavior enables us to comprehend the mutually blocking nature
		of processing and transferring activities,
		and the introduced transformation allows us to calculate
		the total apparent execution times via making simple calculations, essentially (spatial) trigonometry. 
		Our idea
		makes evident, that \textit{having idle time is a natural attribute 
			of technical computing systems}. The inherited technical solutions
		(developed decades ago, under different technological conditions)
		substantially increase the amount of idle time, and the presently 
		experienced stalling is mostly a consequence of the improper technical implementation.
		The idle time of computing systems could be significantly reduced,
		if the technical implementations would consider the timely 
		behavior of the used components appropriately, and they could show the way towards a \textit{more efficient implementation of the "von Neumann architecture"}.
		To reach that goal, however, new design methods must be developed.
		
		The paper discusses the well-established theoretical basis to the necessary depth; furthermore discusses several case studies.
		The listed examples range from simple one-bit adder, through an
		analysis of commonly used technical implementations of serial bus and distributed computing,
		to the temporal behavior of ANNs.
		The temporal behavior also explains, quantitatively, how much performance enhancement
		we can expect from reducing floating operand length, especially under the workload that Artificial Intelligence applications represent. 
		It is also shown that transfer time and processing time
		must be concerted appropriately. It has little sense to miniaturize without
		decreasing the processing time. 	 We demonstrate that replacing some components with a much quicker one
		as well as developing expensive new technologies/materials has a marginal impact
		on the resulting payload performance
		until their access time can be proportionally decreased.
		
	\end{abstract}
	
	\begin{IEEEkeywords}
		computing, technical implementation, modern paradigm, temporal logic,
		computer science
	\end{IEEEkeywords}

	\section{Introduction}
	
	%Computing science is on the border of mathematics and, through its physical implementation,  science.
	Since the beginning of computing, the computing paradigm itself, "\textit{the implicit hardware/software  contract}~\cite{AsanovicParallelCACM:2009}", defined how mathematics-based theory and its science-based implementation must cooperate.
	Computing uses the general operating model, that data is delivered to the input of the processing element, and after some processing time, the resulting data is delivered to some store.  Mathematics, however, considers only the \textit{logical dependencies} 
	between its operands. It assumes that
	the needed operands are instantly available, including that the result becomes an operand for the next operation immediately,
	only after performing the actual operation. 
	
	That is, computing science considers that performing operations, delivering operands
	to and from processing units, is as kind of engineering imperfectness.
	At the time when von Neumann proposed his famous abstraction\footnote{He proposed a structure (functional grouping) of the "computing organs", and NO architecture}, both time of processing
	and time of accessing data (including those on a mass storage devices) were in the millisecond region,
	while the physical data transfer time
	was in the range of microseconds, i.e., three orders of magnitude smaller.
	\textit{It was a plausible assumption to consider that the total time of processing
		comprises only the time of performing the computation plus the time of accessing the required data
		by the operating unit; the physical data delivery time was neglected}.
		
	For today, however, the technical development changed the relations between those timings to their exact opposite. Today the data transfer time (and their access time) is much larger than the time needed to process them.
	Besides, the relative weight of the data transfer time has grown tremendously because of several reasons.
	Firstly, miniaturizing the processors to sub-micron size, while keeping the rest of the needed components (such as buses) above the centimeter scale. Secondly, the single-processor performance stalled~\cite{GameOverYelick:2011}, because they reached the limits,
	the laws of nature enable~\cite{LimitsOfLimits2014}. Thirdly, making truly parallel computers failed~\cite{AsanovicParallelCACM:2009}, and that goal had to be replaced with \textit{parallelized sequential computing},
	disregarding that the operating rules of that different kind of computing~\cite{ScalingParallel:1993}\textbf{\cite{VeghReevaluate:2020,VeghHowMany:2020}} sharply differ from those of the segregated processors.
	Fourthly, the mode of utilization (mainly multitasking) forced out using \gls{OS}s, which imitate a "new processor" for a new task, at serious time expenses.
	Finally, the idea of "real-time connected everything"  introduced geographically vast distances with the corresponding several millisecond data delivery times and assumes that "everything is in cache", which represents a drastically different memory subsystem.
	
	The theory of computing kept the idea of "instant delivery"; although even within the core, wiring has an increasing role~\cite{LimitsOfLimits2014}. Even, since the technology broke through the $0.2~\mu m$ technology bound, the wiring delay dominates the apparent performance~\cite{WiringDominance:2019} (over the gate delay).  The idea of non-temporal behavior, however, was confirmed by accepting "weak scaling"~\cite{Gustafson:1988}, 
	suggesting that \textit{all housekeeping times, such as organizing the joint work of the parallelized serial processors, sharing resources, using exceptions and \gls{OS} services, delivering data between processing units and data storage units, even through slow networks and satellite relay stations, are negligible}.

	Vast computing systems can cope with their tasks with growing difficulty, enormously decreasing computing efficiency, enormously growing energy consumption and heat production.
	Being not aware of that the collaboration between processors 
	needs a different approach (another paradigm),
	resulted 
	in demonstrative failures already known (such as the supercomputers Gyoukou and Aurora'18, or the brain simulator SpiNNaker)\footnote{The explanations are quite different: Gyoukou was withdrawn after its first appearance; Aurora failed: retargeted and delayed; Despite the failure of SpiNNaker1, the SpiNNaker2 is also under construction~\cite{SpiNNaker2:2018};
		"Chinese decision-makers decided to withhold the country’s newest Shuguang supercomputers even though they operate more than 50 percent faster than the best current US machines".} and many more  may  follow:  such as Aurora'21~\cite{DOEAurora:2017},
	the China mystic supercomputers\footnote{https://www.scmp.com/tech/policy/article/3015997/china-has-decided-not-fan-flames-super-computing-rivalry-amid-us} and
	the EU planned supercomputers\footnote{https://ec.europa.eu/newsroom/dae/document.cfm? doc\_id =60156}. 
	Even the successful supercomputers do not develop, and the new (as of June, 2020) supercomputer $Fugaku$ stalled at 
	40\% of its planned capacity.
	
	General-purpose computing systems comprising "only" millions of processors already show the issues, and brain-like systems want to comprise at least four orders of magnitude higher number of computing elements. When targeting neuromorphic features such as "deep learning training", the issues start to manifest at just a couple of dozens of processors~\cite{DeepNeuralNetworkTraining:2016}\textbf{\cite{VeghAIperformance:2020}}. The scaling is nonlinear~\textbf{\cite{VeghReevaluate:2020,VeghScalingANN:2020}}, strongly depending on the workload type, and the \gls{AI}-class workload is one of the worst workloads~\textbf{\cite{VeghAIperformance:2020,VeghScalingANN:2020}} one can run on conventional architectures.\footnote{ https://www.nextplatform.com/2019/10/30/cray-revamps-clusterstor-for-the-exascale-era/ :
		\textit{artificial intelligence, \dots it's the most disruptive workload from an I/O pattern perspective}}

	"\textit{Successfully addressing these challenges [of neuromorphic computing] will lead to a new class of computers and systems architectures}"~\cite{NeuromorphicComputing:2015}. However,
	the roundtable concentrated \textit{only} on finding new materials
	and different gate devices.
	\textit{They did not even mention that for such systems new computing paradigm may also be needed}.   
	The result was that, as noticed by judges of the Gordon
	Bell Prize,  \textit{"surprisingly, [among the winners of the supercomputer competition] there have been no brain-inspired massively parallel specialized computers"}~\cite{GordonBellPrize:2017}.
	Despite the vast need and investments, furthermore the concentrated and coordinated efforts, just because of the important bottleneck:\textbf{ \textit{the missing theory}}.
	
	When thinking about \textbf{Re-Booting} Computing, it is worth to re-read the assumptions made when \textbf{Booting} was made. In his famous publication~\cite{EDVACreport1945}, section 6.3, von Neumann discusses\footnote{As von Neumann formulated: "6.3 At this point the following observation is necessary. In the human nervous system the conduction
		times along the lines (axons) can be longer than the synaptic delays, hence our above procedure of
		neglecting them aside of $\tau$ [the processing time] would be unsound. In the actually intended vacuum tube interpretation,
		however, this procedure is justified: $\tau$ is to be about a microsecond, an electromagnetic impulse
		travels in this time 300 meters, and as the lines are likely to be short compared to this, the conduction
		times may indeed by neglected. (It would take an ultra high frequency device -- $\approx 10^{-8}$ seconds
		or less -- to vitiate this argument.)"~\cite{EDVACreport1945}}
that, \textit{since the processor frequency exceeded 0.1~GHz, it is surely \textit{unsound} to use von Neumann's  computing paradigm in its unchanged form.} For rebooting, the plausible first step is to consider the timely behavior of computing components, including wires. Other questions,
such as which technology, new physical effects/devices, can follow only when the computing paradigm is upgraded and is state-of-the-art.

	\begin{figure*}[t!]
		\centering
		\begin{subfigure}[t]{0.39\textwidth}
			\centering
			\includegraphics%[width=.39\textwidth]
			{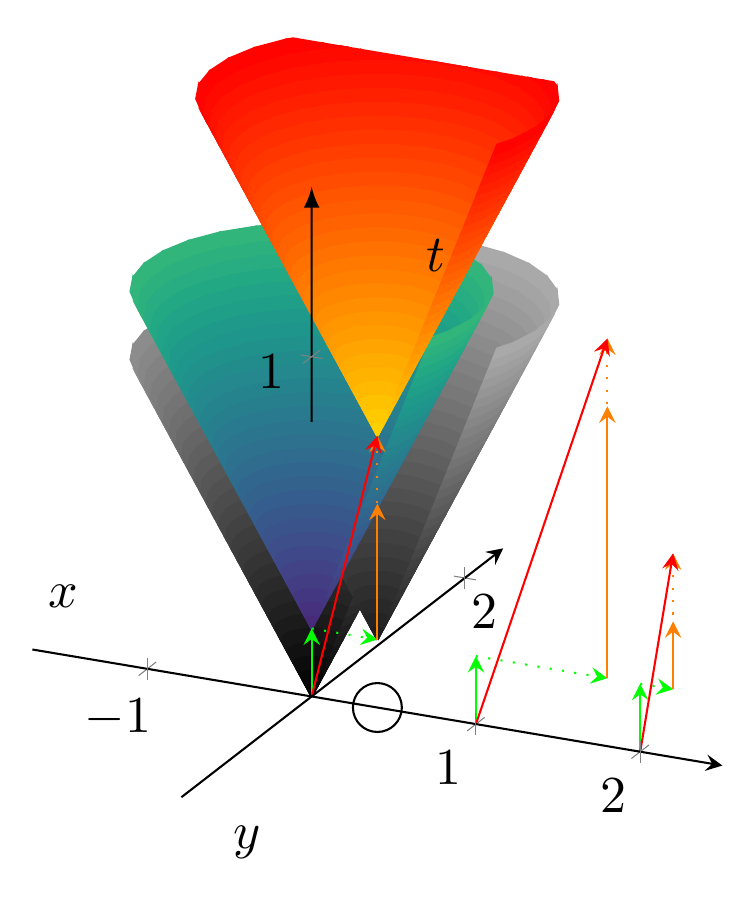}
			\caption{The computing operation in time-space approach. The processing operators can be gates, processors, neurons or networked computers.\label{fig:RelativisticComputation}}
		\end{subfigure}%
		~ 
		\begin{subfigure}[t]{0.61\textwidth}
			\centering
			\includegraphics%[width=.61\textwidth]
			{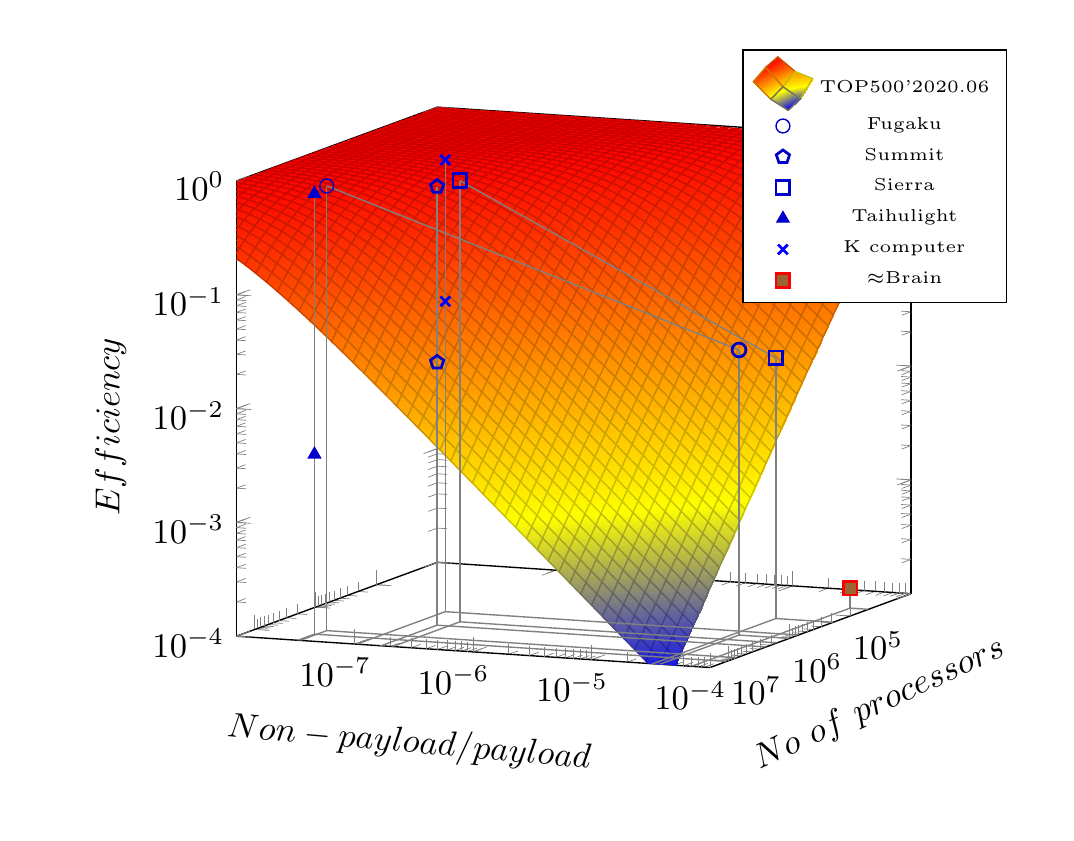}
			\caption{The surface and the figure marks show at what efficiency the top supercomputers run the
				'best workload' benchmark \gls{HPL}, and the 'real-life load' \gls{HPCG}.~\cite{VeghHowMany:2020}
				The right bottom part displays the expected efficiency~\cite{VeghBrainAmdahl:2019} of running
				neuromorphic calculations on \gls{SPA} computers.\label{fig:EfficiencyIdleTime}}
		\end{subfigure}
		\caption{The origin of "idle waiting time" and its effect on the efficiency on parallelized sequential processing systems}
		
		\vspace{-\baselineskip}
	\end{figure*}

	\section{Introducing time to computing}

	As suspected by many experts, the computing paradigm itself%, "\textit{the implicit   hardware/software  contract}~\cite{AsanovicParallelCACM:2009}",
	is responsible for the experienced issues:
	"\textit{No current programming model is able to cope with this development [of processors], though, as they essentially still follow the classical van Neumann model}"~\cite{SoOS:2010}.
	When thinking about "advances beyond 2020", the solution was expected from the "\textit{more efficient implementation of the von Neumann architecture}"~\cite{DeBenedictis_supercomputing:2014}, however.
	Even when speaking about 
	building up computing from scratch ("rebooting the model"~\cite{RebootingComputingModels:2019}), only implementing different gating technology for \textit{the same computing model} is meant. However, the paradigm prevents building large systems (among others, neuromorphic systems), too.
	
	There are many analogies between science and computing~\textbf{\cite{VeghModernParadigm:2019}}; among others, how they handle time.
	Both classic science and classic computing assume instant (infinitely fast) interaction between its objects.
	An event happening at any location
	can be instantly seen at all other locations: the time has no specific role, and an
	event has an immediate effect on all other considered objects.
	In science, inventing that the speed of light is insurmountable, led to introducing the \textit{four-dimensional space-time}.
	Special relativity introduces a 'fourth space dimension', and \textit{we calculate that coordinate of the Minkowski space from the time as the distance the light traverses in a given time}.
	
	To introduce a \textit{temporal logic} into computing, the reverse of that transformation is required. In computing, distances get defined during the fabrication of components and assembling the system. In biological systems, nature defines the neuronal distances, and in 'wet' neuro-biology, signal timing rather than axon length is the right (measurable) parameter. To describe the temporal operation of computing systems correctly, \textit{we need to find out how much later a component notices that an event occurred in the system}.
	That is, we need to use a special four-vector, where all coordinates are time values:
	the first three are the corresponding local coordinates (distances from the location of the event, 
	divided by the speed of the interaction) having time dimension,
	and the fourth coordinate is the time itself; that is,
	we introduce a \textit{four dimensional time-space} system. The resemblance with the Minkowski space is obvious,  and the name difference signals the different aspects of utilization.

	Figure~\ref{fig:RelativisticComputation} (essentially a light cone in 2D space plus a time dimension) shows \textit{why time must be considered explicitly in all kinds of computing}. The figure shows (for visibility) a 3-dimensional coordinate system:
	how an event behaves in a two-dimensional space (the concept is more comfortable to visualize with the number of spatial dimensions reduced from three to two).
	In the figure, the direction 'y' is not used, but enables us to place observers at the same distance
	from the event without locating them in the same point.
	The event happens at the point (0,0,0), the observers are located on the 'x' axis; the vertical
	scale corresponds to the time.
	
	In the classic physical hypothetical experiment, we switch on a light in the origo, and the observer
	switches his light when notices that
	the first light was switched on. 
	If we graph the growing circle around the vertical axis of the graph representing time,
	the result is a cone, known as the \textit{future light cone}.
	Both light sources have some "processing time",
	that passes between noticing the light (receiving the instruction)  and switching the light on (performing the instruction).
	That is, the instruction is received at the origo, at the bottom of the green arrow.
	The light goes on at the head of the arrow (i.e., at the same location, but at a later time),
	after the 'processing time' $T_p$ passed. Following that, the light propagates
	in the two spatial dimensions as a circle around axis "t".
	Observers at a larger distance notice the light at a later time:
	a 'transmission time' $T_t$ is needed.
	If the "processing time" of the light source of the first event were zero,
	the light would propagate along the gray surface at the origo.
	However, because of the finite processing time, the light will propagate along the
	blueish cone surface, at the head of the green arrow.

	\begin{figure*}
		\includegraphics[width=\textwidth]{%fig/
			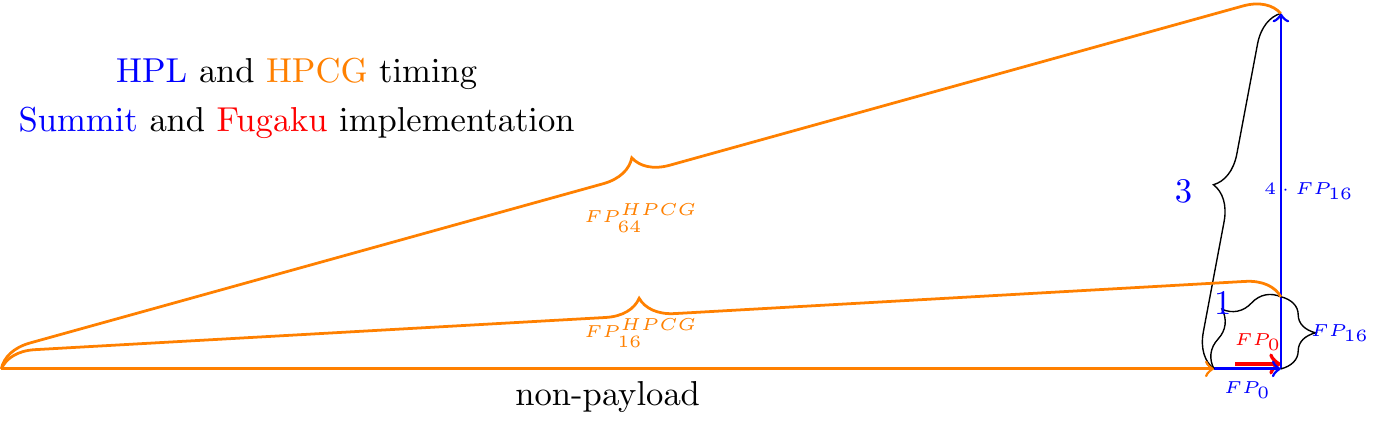}
		\caption{The role of non-payload contribution in defining $HPL$ and $HPCG$ efficiencies,
			for double and half precision floating operation modes, using different architectural concepts. For visibility, a hypothetic efficiency ratio $E_{HPL}/E_{HPCG}$=10 assumed. The housekeeping (including transfer time and cache misses)
			dominates, the length of operands has only marginal importance.\label{fig:FP16vsFP64HPCG}}
	\end{figure*}

	A circle denotes the position of our observer on the axis "x".
	With zero "transmission time", the second gray conical surface (at the head of the green dotted arrow)
	would describe his light. However, its "processing time" can only begin when the observer notices the light at his position: when the dotted red arrow hits the blueish surface.
	At that point begins the "processing time" of the second light source;
	the yellowish conical surface describes the second light propagation.
	The horizontal (green dotted) arrow describes the physical distance of the observer (as a time coordinate),
	the vertical (red dotted) arrow describes the time delay of the observer light.
	It comprises two components: the $T_t$ transmission time to the observer and its $T_p$ processing time. The light cone of the observer starts at $t=2*T_p+T_t$.
	
	The red arrow represents the resulting \textit{apparent processing time} $T_A$: 
	the longer is the red vector; the slower is the system.
	As the vectors are in the same plane,
	$T_A = \sqrt{T_t^2+(2\cdot T_p+T_t)^2}$, that is  $T_A = T_p\cdot \sqrt{R^2+(2+ R)^2}$.
	This means, that \textit{the apparent time is a non-linear function 
		of both of its component times} and \textit{their ratio $R$}.
	If more computing elements are involved, $T_t$ 
	denotes the longest transmission time. (Similar statement is valid if the $T_p$ times are different) The effect is significant:  if $R=1$, the apparent execution time of performing the two computations is more than 3 times longer than the processing time.
	Two more observers are located on the axis 'x', in the same position.  For visibility, their timings are displayed at points '1' and '2', respectively.  Their results illustrate the influence of the transmission speed (and/or the ratio $R$).
	In their case \textit{the transmission speed differs by a factor of two} compared to that displayed at point '0'; in this way,
	three different $R=T_t / T_p$ ratios are displayed.
	
	Notice that at half transmission speed (the horizontal green arrow is twice as long as that in the origo)
	the vector is considerably longer, while at
	double transmission speed, the decrease of the time
	is much less expressed\footnote{\cite{VeghHowMany:2020} discusses this phenomenon in detail.}.

	\section{Idle times in computing}
	Specialized standard (benchmarking) programs are used to characterize the payload performance of a computing system. Those benchmarks programs express the performance of the system in such a way that the known number of the executed "payload" instructions are
	divided by the total measurement time.
	In this way, all activities, made by the computing system, are included in the total measurement time.
	The "idle time" is present in the measurement,
	but given that the classic computing assumes instant interaction, it is assumed to be zero.
	Because of this, it is commonly believed that the housekeeping activity (including looping,
	addressing, branching) is negligible (or maybe it
	can be done, at least partly,  in parallel with the payload computing).
	However, the calculation cannot begin until the operand
	reaches the input of the processing unit, and similarly, the result
	cannot be transferred until the computation finished:
	the transfer and computation operations mutually
	block each other.
	
	From the point of view of
	computation, $T_p$ is a "payload time", and 
	$T_t$ is a "non-payload (although absolutely necessary) time".
	Only the "payload computing" contributes to
	"payload time". The time needed to deliver
	the data from one place to another, with the limiting speed of interaction, contributes to the "non-payload" time, see their projection to the time axis. That is, even single-processor performance
	measurements include some amount of idle time.
	The measurable single-processor payload performance depends on the underlying architectures and the measured instruction mix (the benchmark program).
	The different benchmarks represent a different instruction mix, to which different effects, such as cache utilization, are attached (and so: they represent a different workload). In this way, the same computing systems show 
	different payload performances when measured by
	different benchmarks~\cite{DifferentBenchmarks:2017},~\textbf{\cite{VeghHowMany:2020}}. This difference gets remarkable,
	when benchmarking systems, comprising parallelized single processors.
	
	In our simple case, payload and non-payload times are interpreted for single computing events, but the same
	interpretation is valid for the average transfer times  $T_A$.
	The blocking nature of transfer persists.
	Under different conditions, say, when the average number of cache failures changes, the apparent execution time 
	$T_A$ (and so: the apparent computing performance) changes.
	That is, an idle time naturally belongs to a computing 
	operation, and its ratio to the payload operation, defines the computing efficiency of the computing system.
	
	Notice one more important aspect:
	\textit{the $T_p$ transmission time is an 'idle time'} (the orange arrow on the figure) for the observer:
	it is ready to run, takes power, but does no useful work.
	Due to their finite physical size and limited interaction speed (both neglected in the classic paradigm), \textit{temporal operation of computing systems 
		results inherently in an idle time of their
		processing units\footnote{it can be a crucial factor of inefficiency of general-purpose chips~\cite{InefficiencyHameed2010}}}, and -- since it sensitively depends on many factors and conditions -- \textit{can be a significant contributor to
		non-payload portion of their processing time}. 
	Other major contributors originate from their technical implementation of the components, and these "idle waiting" times sharply decrease the payload performance of the systems. Figure~\ref{fig:EfficiencyIdleTime} depicts
	how the efficiencies of recent supercomputers depend~\textbf{\cite{VeghHowMany:2020}} on the number of single-threaded processors in the system and the
	parameter $(1-\alpha)$, describing the non-payload portion of the corresponding benchmark task. It is known for decades that "\textit{this decay in performance is not a fault of the
		architecture, but is dictated by the limited parallelism}"~\cite{ScalingParallel:1993}; in excessive systems of modern \gls{HW}, \textit{is also dictated by the laws of nature}~\textbf{\cite{VeghModernParadigm:2019}}. 
	
	\begin{figure*}[t!]
		\centering
		\begin{subfigure}[t]{0.5\textwidth}
			\hspace{-2cm}
			\includegraphics[width=1.2\textwidth]
			{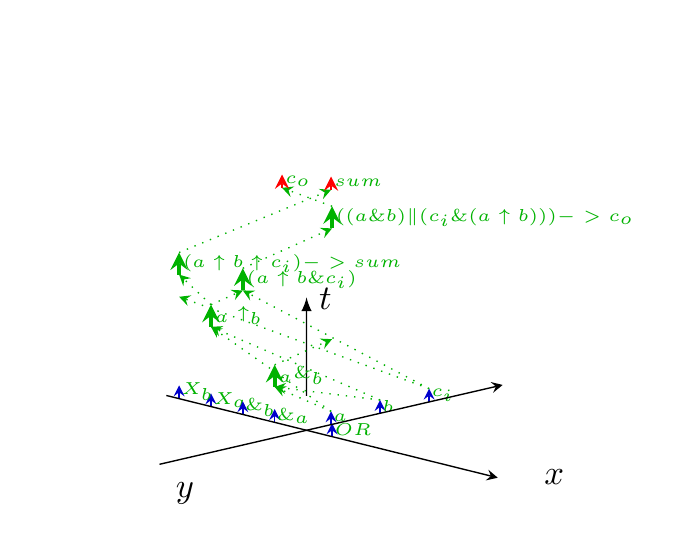}
			\caption{Temporal dependence diagram of a 1-bit adder. The second XOR gate is at (-1,0).
				\label{fig:Adder1bit}
			}
		\end{subfigure}%
		~ 
		\begin{subfigure}[t]{0.5\textwidth}
			\hspace{-2cm}
			\centering
			\includegraphics[width=1.2\textwidth]
			{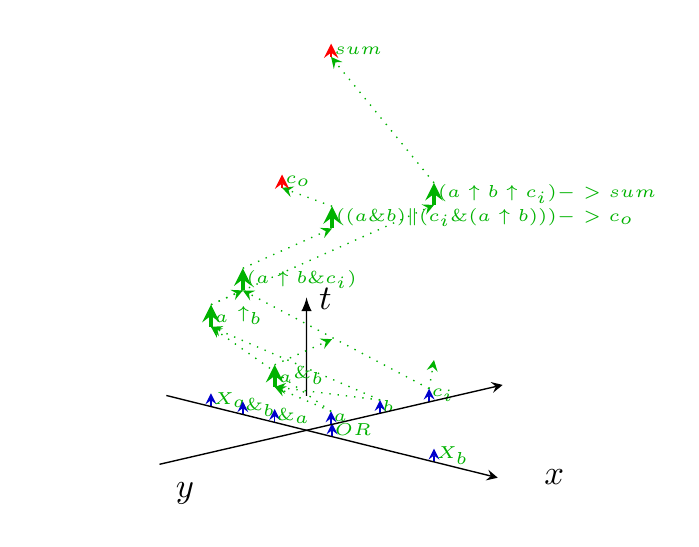}
			\caption{Temporal dependence diagram of a 1-bit adder. The second XOR gate is at (+1,0).			
				\label{fig:Adder1bitPlus}
			}
		\end{subfigure}
		\caption{Temporal diagram of a one-bit adder in time-space system. The diagram shows the logical equivalent of the SystemC source code of Listing~\ref{lst:OneBitAdder}.
			\textit{The time from axis $x$ to the bottom of green arrows} signals "idle waiting" time (undefined gate output). Notice how changing the position of a gate affects signal timing.\label{fig:OneBitAdder}}
		%	\vspace{-\baselineskip}
	\end{figure*}
	
	\subsection{Half versus double precision}
	One of the most clean conditions to demonstrate
	the impact of  idle time on the computing
	performance of the system, is to use different operand
	lengths in the computing operations, but otherwise using
	the same underlying architecture, number of operations, and housekeeping.
	From our point of view, only the amounts of payload and non-payload operations are different.
	
	The so-called \textit{HPL-AI} benchmark used Mixed Precision\footnote{Both names are rather inconsequent. On one side, the test itself has not much to do with AI, just uses the operand length common in \gls{AI} tasks (\gls{HPL}, similarly to \gls{AI}, is a workload type). On the other side, Mixed Precision is Half Precision: it is natural that for multiplication twice as long operands are used temporarily.  It is a different question that the operations are contracted.}
	\cite{MixedPrecisionHPL:2018} rather than Double Precision 
	calculations. Recent top supercomputers  $Fugaku$~\cite{DongarraFugakuSystem:2020} and $Summit$~\cite{MixedPrecisionHPL:2018} provided also their \gls{HPL} performance 
	for both 64-bit and 16-bit operands. Of course, the performance seems to be much better with shorter operand length
	(much shorter time for the same number of operations).
	One can expect that their performance shall be four times higher, using four times shorter operands.
	The power consumption data~\cite{MixedPrecisionHPL:2018} underpin the expectations:
	the power consumption is about four-fold lower.
	The computing performance, however, shows a slighter performance enhancement only: 3.01 for $Summit$,
	3.42 for $Fugaku$, because of the needed housekeeping. 
	
	Unfortunately, \textit{the achievement comes from accessing less data in memory and using quicker operations on the  shorter operands
		rather than reducing the communication intensity}.
	In other words, it reduces proportionally only the payload performance, but much less the non-payload performance.
	As discussed in detail in~\textbf{\textbf{~\cite{VeghHowMany:2020}}},
	benchmark \gls{HPL} is computing-bound, so the longer (averaged) data delivery time
	blocks the operation of the floating point unit. 
	As seen in the figure, under \gls{HPL} workload, the idle time is smaller than the
	computing time, so using shorter operands leads to a significant difference
	in the payload computing performance of the system.
	In a different workload, where the non-payload portion is much higher, the non-payload contribution
	dominates; in the case of \gls{HPCG} workload, the difference between the lengths of the operands becomes marginal.

	Fig.~\ref{fig:FP16vsFP64HPCG} illustrates the role of non-payload performance with respect to operand length.
	In the figure (for visibility) a hypothetic ratio 10 of efficiencies measured by benchmarks $HPL$ and $HPCG$
	was assumed; in reality, it is about 200. The non-payload contribution (the data transfer) blocks the operation of the floating-point unit and defines the system's payload performance.
	The dominant role of the non-payload contribution also means that 
	it is of little importance if double or half precision operands are used in the computation in real-life tasks.
	The blue vectors essentially represent the case of $Summit$. The red vector represents
	the $FP_0$ value of $Fugaku$ (transformed to scaling of $Summit$).
	The difference between their $HPL$ performances is attributed to their different
	$FP_0$ values. This difference, however, gets marginal as the workload approaches real-life conditions.
	Under \gls{AI} workload, this effect is even more substantial: for \gls{ANN} architectures and workloads,
	using shorter operands means no change in their payload performance.
	
	\subsection{Different workloads}
	Under different workloads a different amount of non-payload contribution is
	attached to the payload contribution, so the resulting apparent payload performance is different. 
	What is more, the efficiency (and so: the payload performance)  depends also
	on the number of processing units in the system~\textbf{\cite{VeghScalingANN:2020}}.
	The efficiency sharply decreases with the growth of the number of processing units,
	so some supercomputers use all their available cores when running benchmark \gls{HPL},
	but only a fraction of them when running \gls{HPCG}. Even, some of them (typically cloud-like architectures) do not provide 
	their \gls{HPCG} efficiency. Figure~\ref{fig:EfficiencyIdleTime} reveals how supercomputers' efficiencies depend on their number of processors and the workload they run. Moreover, it also demonstrates why is more advantageous to use only a fraction of their available cores (compare the efficiency values for $Fugaku$ and $Taihulight$).

	\begin{lstlisting}[float,caption=The essential lines of source code of the one-bit adder implemented in 
	SystemC,label=lst:OneBitAdder]
	//We are making a 1-bit addition
	aANDb = a.read() & b.read();
	aXORb = a.read() ^ b.read();
	cinANDaXORb = cin.read() & aXORb;
	
	//Calculate sum and carry out
	sum = aXORb ^ cin.read();
	cout = aANDb | cinANDaXORb;
	\end{lstlisting}
	\subsection{Programing as a source of idle time}
	Fig.~\ref{fig:RelativisticComputation} shows how the physical implementation (the finite size of computing components) introduces idle time into computing. The "von Neumann-style" programming~\cite{BackusNeumannProgrammingStyle} also adds its contribution. Although the corresponding code shown in Listing~\ref{lst:OneBitAdder} assumes  only logical dependence between the
	gate operations, their physical implementation converts that
	logical dependence to a temporal one, as shown in Fig.~\ref{fig:OneBitAdder} for the case of a one-bit adder. Changing 
	the \textit{position} of one single gate changes the operating time of the gate.

\begin{figure*}
	\centering
	\hspace{-.2cm}\begin{subfigure}[t]{.55\textwidth}
		\centering
		\includegraphics[width=1\textwidth]
		{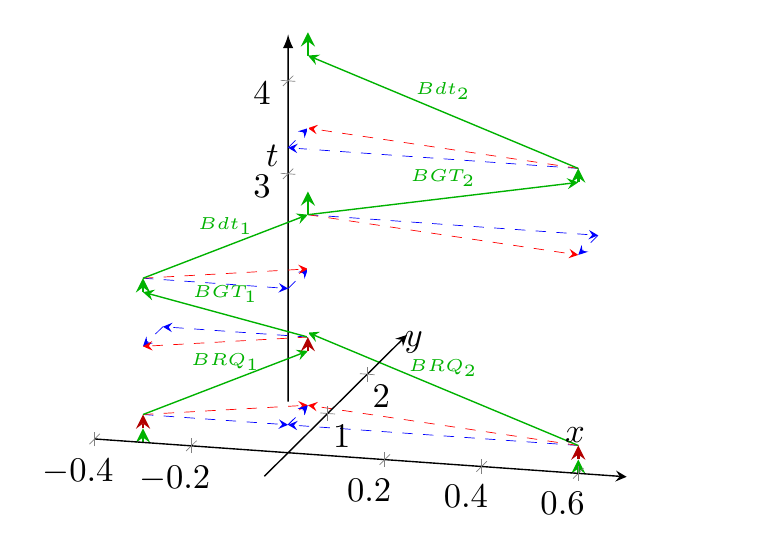}
		\caption{The operation of the sequential bus, in time-space coordinate system system. Near to axis \textit{t}, \textit{the lack of vertical arrows} signals "idle waiting" time\label{fig:Relativisticbus}}
	\end{subfigure}%
	\hspace{.25cm}\begin{subfigure}[t]{.43\textwidth}
		\centering
		\includegraphics[width=1\textwidth]
		{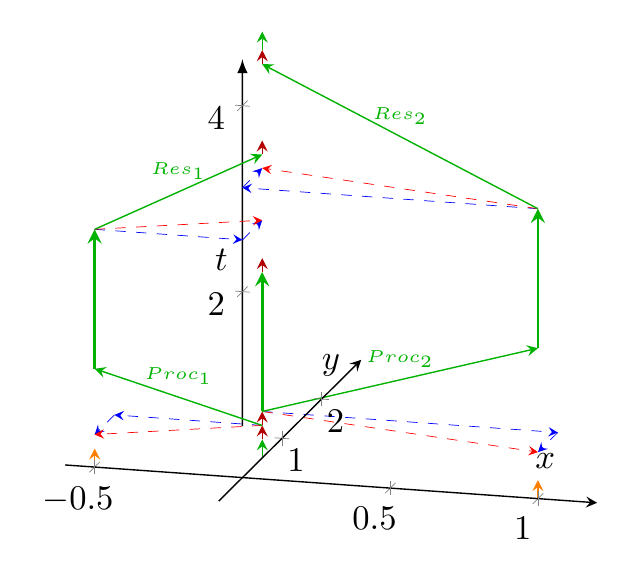}
		\caption{The parallelized sequential operation as described in the proposed time-space system
			\label{RelativisticDistributed}}
	\end{subfigure}
	%\vspace{-\baselineskip}
	\caption{The operation of two popular technical implementations in the time-space coordinate system \label{fig:TechnicalImplementations}}	
	\vspace{-\baselineskip} 
\end{figure*}

	\subsection{Technical implementations}
	The idle time, the effect of which is depicted in Fig.~\ref{fig:EfficiencyIdleTime},
	is "including, but not limited to," the idle times shown in Fig.~\ref{fig:RelativisticComputation}. In parallel with the computing paradigm,
	its technical implementations have also been inherited from the former solutions.
	Fig.~\ref{fig:Relativisticbus} depicts a common implementation of a 
	"high-speed bus", connecting computing components. 
	The fundamental issue with using a shared medium is that -- for the time of the 
	actual transfer --  the bus must be private, i.e., the components connected to the bus
	must compete for the right of "owing" the medium.
	As displayed, the overwhelming majority of the operating time is spent with
	competing for the bus, and the advantage of the "high speed" can be used
	in the little portion of the total time. In the case of \gls{ANN}s, 
	all neurons (in the layer) want to use that shared bus at the same time.
	In this way, the role of the high speed becomes marginal, as discussed in~\textbf{\cite{VeghScalingANN:2020}}.
	
	Fig.~\ref{RelativisticDistributed} displays the temporal diagram of
	distributed (parallelized sequential) processing. The idea that
	one of the processing units organizes the joint work works fine only
	until the number of the joint processing units is relatively low.
	As the number of fellow processing units grows, so increases the idle time of the organizing unit.
	For a detailed (and quantitative) discussion see~\textbf{\cite{VeghHowMany:2020}}.

	\subsection{Impact of temporal behavior on ANNs}
	The technical implementation may have 
	vital impact on the operation of \gls{ANN}s.
	The results of the neuronal calculations
	are serialized, and indefinitely delayed, 
	as discussed in~\textbf{\cite{VeghScalingANN:2020}}.
	The good solution,
	the biology applies, that all results are 
	sent on the corresponding axons in parallel, that is, at the same time (or at least within a well-defined time window).
	A proper technical solution could be to make sure that all
	needed neuronal operations are performed, that would need a synchronized operation.
	However, that solution would slow down the operation.
	The present (wrong) technical solution is that
	the vector and matrix operations are performed on their data without synchronization, as fast as the HW can perform the operations. The processed data is derived from the outputs of
	artificial neurons. Unlike their biological equivalents, those technical neurons produce "continuous spikes", and there is no way to find out which input signals were used to compute that output signal. During training, the computations' results are continuously used to make feedback, i.e., to adjust the weights, the neuronal calculations use.
	At the beginning of the computations, 
	those vectors are not yet initialized:
	their correct input value is waiting in the queue. The calculation with those uninitialized vectors (and matrices) are performed,
	and the feedback adjusts the weights to those fake inputs. This temporal behavior makes it significantly harder to learn something,
	because the weights may be adjusted tendentiously to wong sets.
	Ironically enough, the more neurons and
	the faster (maybe, tensorial) math units, 
	the harder is to arrive at a reasonable solution: not considering the temporal behavior of their implementation,
	misleads the neuronal networks and leads to weeks-long training times.
	
	A further issue is discussed in detail in~\textbf{\cite{BiologySpatioTemporal:2020}}: the time is a crucial attribute in learning (and training). Given that in artificial networks the time as such is not considered at all (both in math discussion and technical implementation), the produced behavior is practically unrelated to the biological systems they attempt to mimic.
	
\begin{figure*}
	\centering
	\hspace{-1cm}\begin{subfigure}[t]{.45\textwidth}
		\centering
		\includegraphics[width=1.2\textwidth]
		{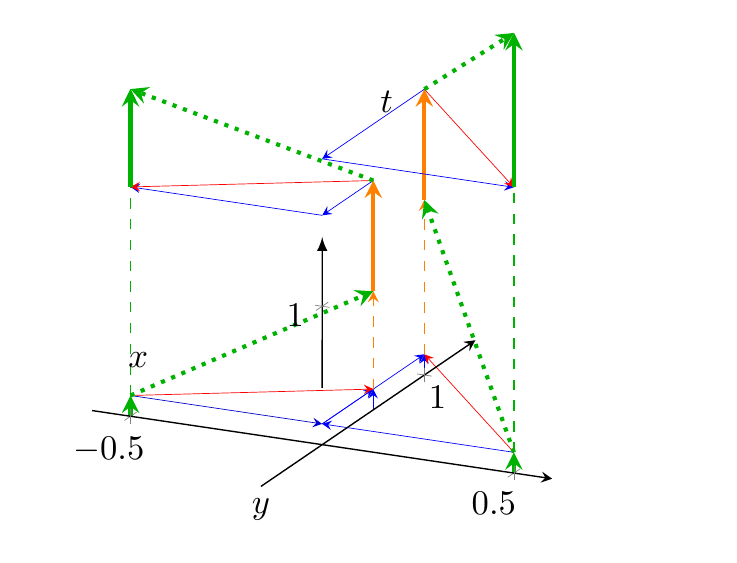}
		\caption{Normal speed cache memory. Two different cache memories, with the same physical cache sped, but at different internal on-chip cache position\label{fig:RelativisticMemorySlow}}
	\end{subfigure}%
	\hspace{.5cm}\begin{subfigure}[t]{.45\textwidth}
		\centering
		\includegraphics[width=1.2\textwidth]
		{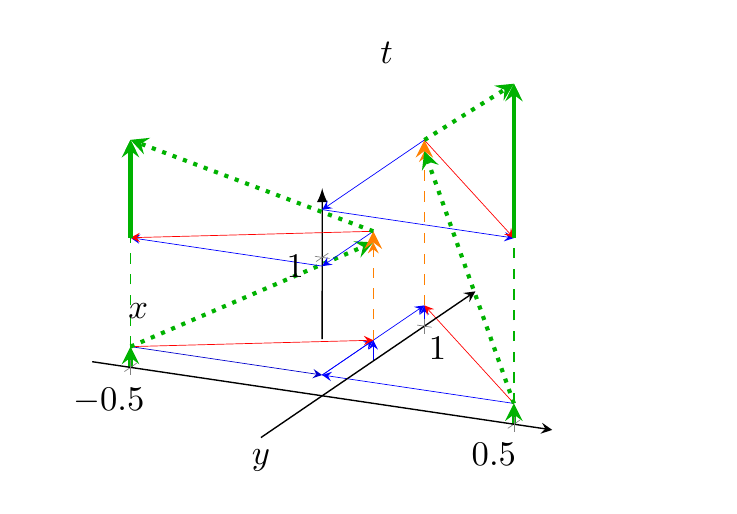}
		\caption{Super-quick (10 times quicker) cache memory. Assumes new material/physical mechanism. Two different cache memories, with the same physical cache sped, but at different internal on-chip cache position \label{fig:RelativisticMemorySuper}}
	\end{subfigure}
	%\vspace{-\baselineskip}
	\caption{The performance dependence of an on-chip cache memory, at different cache operating times, in the same topology.
		The cores at x=-0.5 and x=0.5 positions access on-chip cache at y=0.5 and y=1.0, respectively. The vertical orange arrows represent the physical cache operating time, and the vertical green arrows the apparent access time. The physical operating speed of the cache memory of the right subfigure is ten times better. Compare the apparent access times to the corresponding physical ones: the time ratio is better only about a factor of two. Notice also that the apparent operating speed is more sensitive to the position rather than to the speed of the cache memory\label{fig:CachePerformance}}	
	\vspace{-\baselineskip} 
\end{figure*}

	\subsection{New effects and/or materials}
	Given that the \textit{apparent processing time} $T_A$ defines the performance of the system, $T_p$ (physical processing time,  a vector perpendicular to the XY plane) and $T_t$ (transfer time, a vector between different planes) must be concerted.
	Their temporal behavior defines the limitations of researching
	for new materials/effects, etc.
	In a complex system, \textit{it is not reasonable to fabricate smaller components without decreasing 
		their processing time proportionally; and similarly, replacing a \gls{PU} with a very much quicker one
		(such as proposed in~\cite{RecipeMemristor:2020,NatureBuildingBrain:2020}, and \textit{may be proposed using any future new physical effect and/or material}) 
		has only a marginal effect, if the physical distance
		of the \gls{PU}s cannot be reduced proportionally, at the same time.}

	Fig.~\ref{fig:CachePerformance} demonstrates why: two different topologies and two different 
	physical cache operating speeds are used in the figure. 
	Two cores are in positions (-0.5,0) and (0.5,0), furthermore two cache memories are located at (0,0.5) and (0,1). 
	The signal, requesting to access cache, propagates along the dotted green vector
	(it changes both its time and position coordinates), the cache starts to operate only when the green dotted arrow hits its position. After its operating time (the vertical orange arrow), the result is delivered back to the requesting core.
	This time can also be projected back to the "position axes", and their sum (thin red arrow) can be calculated.
	
	The physical delivery of the fetched value begins at the bottom of the lower vertical green arrows, includes waiting %(dashed thin green line)
	and finishes at the head of
	the upper vertical green arrows; their distance defines the \textit{apparent cache access time} $T_A$. The physical cache access time (the vertical orange arrow) begins when the
	signal reaches the cache. Till that time, the cache is idle waiting. 
	The core is also idle waiting until the requested content arrives.
	Notice that \textit{the apparent processing time is a monotonic function of the
		physical processing time, but because of the included -- fixed time -- 'transmission times'
		due to the physical distance of the respective elements, their dependence is far from being linear}.
	Repeated operation of course can change the idle/to active ratio. 
	However, one must consider the resources the signal delivery uses and the blocking effect discussed above.
	
	The apparent processing time (represented by the distance of the vertical green arrows) is only slightly affected by the physical speed of the cache memory (represented by vertical orange arrows).
	The right subfigure assumes that some new material/technology/effect decreased the
	access time to one-tenth of the time assumed on the left subfigure.
	In the figure, the technology (at considerable expenses) improved the physical access time by a factor of ten, but the apparent access speed has improved only by a factor of less than two.
	\textit{Even if the physical cache time could be reduced to zero, the apparent access time cannot be
		reduced below the time defined by the respective distances/interaction times.}
	As discussed theoretically in~\cite{LimitsOfLimits2014} and in terms of practical design~\cite{WiringDominance:2019}, in the today's processor technology wiring defines performance.
	
	\section{Summary}
	All fields of computing benefit from introducing temporal behavior for the components,
	from explaining the need of "in-memory computing" to reasoning the low efficiency of \gls{GPU}s in general-purpose applications. We presented some possible case studies.
	\textit{Neglecting their temporal behavior limits the utility of any new method, component, material or technology, if they are designed/developed/used in the spirit of the old (timeless) paradigm.}
	As Neumann pointed out, it is \textit{\textbf{unsound}} to establish
		supercomputing, different accelerators, ANNs, memristor, FPGA, quantum-processor, etc. based computing on the classic paradigm that is valid (in a quasi-strict mathematical sense) for the technological age of vacuum tubes.
		The drastically increased weight of the transfer time compared to the computing time needs a modern paradigm.
	
%	\bibliographystyle{IEEEtran}
%	%		\bibliography{CSCEtime}
%	
%	\bibliography{../../CommonBibliography%
%		,../../CommonPrivateBibliography%
%		,../../CommonNeuronalBibliography%

\end{document}